\documentclass[twocolumn,aps,pre,showpacs,preprintnumbers,amsmath,amssymb]{revtex4-1}
\usepackage{amsmath}
\usepackage{mathtools}
\usepackage{amsfonts}
\usepackage{amssymb}\usepackage{graphicx}
\usepackage{color}
\usepackage[normalem]{ulem}

\usepackage{amssymb,latexsym,mathrsfs}
\newcommand{\be}{\begin{equation}}
\newcommand{\ee} {\end{equation}}

\newcommand{\ra}{\rangle}
\newcommand{\la}{\langle}

\def\bea{\begin{eqnarray}}
\def\eea{\end{eqnarray}}
\def\ba{\begin{array}}
\def\ea{\end{array}}
\def\bm{\begin{matrix}}
\def\em{\end{matrix}}
\def\n{\nonumber} 
\def\c{\mathscr}
 \def\c{\mathscr}

\newcommand*{\bfrac}[2]{\genfrac{\lbrace}{\rbrace}{0pt}{}{#1}{#2}}
\newcommand{\threevec}[1]{\ensuremath{\begin{psmallmatrix} \end{psmallmatrix}}}

\begin{document}
\title{Multi-chain   models  of Conserved Lattice Gas}
\author{Arijit Chatterjee} \email{arijit.chatterjee@saha.ac.in}
\author{P. K. Mohanty} \email{pk.mohanty@saha.ac.in}
\affiliation{CMP Division,  Saha Institute of Nuclear Physics, HBNI, 1/AF Bidhan Nagar, Kolkata 700064, INDIA}  

\begin{abstract}
 
Conserved lattice  gas (CLG) models in one dimension  exhibit absorbing state phase transition (APT) with  simple integer exponents $\beta=1=\nu=\eta$ whereas the same   on a  ladder    belong to  directed percolation (DP)universality. We conjecture  that additional   stochasticity  in  particle transfer  is a  relevant perturbation  and  its  presence 
on a ladder force the  APT   to be in DP class.  To substantiate    this we  introduce  a   class  of 
restricted   conserved lattice gas models  on a  multi-chain system ($M\times L$ square lattice with periodic boundary condition 
in  both directions), where  particles  which   have  {\it exactly one vacant  
neighbor}  are active   and they move  deterministically  to  the   neighboring vacant site.  We show that  for  odd number 
of chains ,  in the thermodynamic  limit $L\to \infty,$  these   models  exhibit  APT  at $\rho_c=  \frac{1}{2}(1+\frac1M)$   
with $\beta =1.$  On the other hand, for  even-chain systems   transition occurs at $\rho_c=\frac12$  with $\beta=1,2$ 
for $M=2,4$ respectively,  and $\beta= 3$ for $M\ge6.$ We illustrate this   unusual critical behaviour analytically 
using a  transfer matrix  method.  
\end{abstract}

\maketitle

\section{Introduction}
In the study of absorbing state phase transition (APT) \cite{Hinrichsen}, directed percolation (DP) \cite{DPRev} 
has been considered to be the most robust universality class. Critical behavior encountered in many diverse problems,  like synchronization\cite{synchro}, damage spreading \cite{damage}, depinning transition \cite{depin},  catalytic reactions \cite{catalytic}, forest  fire \cite{fire},  
extinction of species \cite{bioEvol}  etc.  belong to the DP universality class \cite{DPExp}. It has been conjectured  \cite{DPConjecture}  that in absence  of any special symmetry or quenched randomness,  APT  in systems following short ranged dynamics,
characterized by  a non negative fluctuating  scalar  order parameter, belongs to DP. 
Presence  of additional   conservation laws,   like  particle-hole symmetry \cite{A9}, 
conservation of parity \cite{A10}, and symmetry between different absorbing states \cite{A11} 
lead to different universalities.  Non-DP  behaviour  has also  been   reported  
in sandpile models \cite{A12}   where   the order parameter  itself  does not  have any additional symmetry   but it is  coupled to a   conserved density  or  height field \cite{A13}. 
In fact,   existence of conserved density field is not the sufficient criteria to characterize various universality classes; the noise in the order parameter field due to the dynamics plays crucial role.
There are many examples of systems belonging to DP universality class in   presence of a conserved field,  most important  being the conserved Manna models \cite{B13,assist1,B14, B15,GDP}. These models has recently been claimed   
to belong to the DP class \cite{B16} contrary to  the common  belief that   they   exhibit non-DP critical behaviour. Sticky sand-piles are  another  generic class  of models \cite{B17}  which  show    DP  behaviour in presence of conserved fields. 

APT in presence of a conserved field \cite{CLG,B11} has  been a  subject of interest 
in recent years.  The conserved lattice gas (CLG) model \cite{B12,UrnaPK} and some  of its extensions  \cite{rdd, arijit_pk} are  exactly solvable   in  one dimension and they  
provide  clear    examples of  non-DP   behaviour.  These models are rather 
simple   having  trivial integer exponents. Some  variations of   CLG models also show continuously varying critical exponents or  
multi-critical behaviour \cite{rdd, arijit_pk}.   Non-DP  behaviour in these  models can not be  
blamed to presence of the conserved density  because   the same dynamics  on 
a ladder  geometry lead to an absorbing transition belonging to DP class \cite{B16}. 
In this article we propose that  the   CLG  in  1D  (1DCLG)  belongs to an universality class 
different   from DP   as the  particle transfer  occurs   there  deterministically. We 
show that  if the  dynamics is restricted so that particles  hop  deterministically
the  CLG models  on a ladder  belong to the  universality class  of 1DCLG. A natural question  
is  then, what  is the nature of the absorbing  transition in multi-chain systems ?    

The multi-chain systems introduced in  this article  can be solved  using a  transfer matrix method
by expressing  the steady state  weights as  the trace  of product matrices formed   by   replacing  each rung  by a representative  matrix.  When  the  number of chains  $M$  is an odd integer,   the system  exhibit  an    APT  at   density $\frac12(1 + \frac1{M})$  belonging to  1DCLG universality class  with    the  order parameter exponent $\beta=1.$  On the   other hand,  for even  number  of chains    critical density  turns out to be $\frac 1 2$      and   the order parameter  exponent for   large  $M>4$ is $\beta=3,$ with   an unusual finite size effect for small $M:$ $\beta=1,2$  for  $M=2,4$ respectively.

The   article   is organized as follows. In section  \ref{sec:model}  we  introduce  the   restricted  
CLG dynamics    and  study  APT  in these models   on a ladder  geometry.  Here  we introduce the 
transfer matrix formalism  and obtain the critical exponents   $\beta,  \nu, \eta.$ In sections \ref{sec:Multichain} we  generalize the model for  $M>2$  and   study  the  odd  and even $M$ chains in separate sub-sections.  Finally  in next section we conclude  and discuss    some  important  issues   of the   multi-chain systems  of   conserved lattice gas.

\section{The Model}\label{sec:model}
The   conserved lattice gas model in one dimension \cite{CLG,B12,UrnaPK}  is defined by the dynamics, 
\be
110\to 101 ;~ 011 \to 101.  
\ee
The   dynamics  conserves the number of particle $N$  or the density $\rho= \frac NL.$
The  first part of the  dynamics $110\to 101,$  which   corresponds to  rightward hopping,     
is effectively  a  combination of   $1100\to 1010$ and $1101\to 1011,$ of which the   the former one 
destroys  the    consecutive  zeros (00s)  and consecxutive ones (11s)  if  present in the system and 
the second part is  00 and 11-conserving.  The same    is true  for  left hop  $011 \to 101.$ 
Thus, the  number of consecutive zeros  (CZs)  can only reduce  as  the system evolves. 
Once  the system leave  a configuration  with higher number of  CZs  to  another  with 
lower  number, it  never  visits  it again;  these   non-recurring configurations  
must be absent in the  steady state when  $\rho\ge\frac12$, wheras for 
$\rho\le\frac12$ non  of the configurations  are  devoid of CZs and  the system falls  into an an 
absorbing  configutaion where   consecutive 1s are aslo absent.  From  the exact results \cite{UrnaPK} 
one knows that  the  absorbing transition takes place at   $\rho_c= \frac12$ with critical exponents 
$\beta=1=\nu=\eta.$

To generalize the model  to a  ladder  and  mlti-chain   system,  we notice that  the 
dynamics of one  dimensional  CLG model (1DCLG)   can be interpreted  in  two ways, (a) particles  having 
one  occupied neighbor are  active  and they move   to the   neighboring vacant site  with  unit rate,  
or (b) particles which has  {\it exactly} one vacant  neighbor   move  to {\it that} vacant site. 
A  natural extension  of  (a) to  a  two-chain system (a ladder) would result in a   dynamics,
where  particles    having   atleast  one occupied neighbor   are active  and they   move to  one 
of the  available  vacant  site.  This  dynamics is stochastic, as  each site  on a ladder   has  
three   nearest neighbors and  an active particle may have more than one vacant nearest neighbors 
where it  must   choose  one of them  randomly and independently and hops to that site. 
This  model   was  studied  in   \cite{B16},  which  showed  that  CLG  on a ladder exhibit an absorbing 
phase transition belong to DP-universality class.  Interpretation (b)  can  also  be extended to a 
two-chain  CLG  model, where the dynamics  would  be  deterministic; this is because, now  each active particles 
has  {\it exactly  one vacant neighbor}  it hops  deterministically to that site.   
In  the following we   study   this  dynamics  and  show that  with  this  deterministic dynamics
CLG  on a ladder  belong to  the   universality class of  1DCLG,    with exponents  $\beta=1=\nu=\eta.$ 
It is   no surprise  that   in absence  of stochasticity the  phase transition  different from   
the  most  robust  universality, namely  DP class. However, we observe that these models   on a  multi-chain 
system show many interesting   feature, which we   discuss in  next section.  
First   study the  model for two-chain system   and show  that   this  quasai-1D system  
with deterministic  dynamics are  no different from their one dimensional counterpart.

\subsection{ CLG model on a ladder   with   deterministic dynamics}
\label{sec:IIA}
\begin{figure}
\begin{center}
\includegraphics[width=8.5 cm]{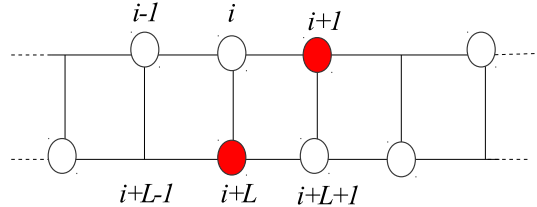}\vspace*{-1cm}
\caption{(Color online) Schematic description of the model: Particles having two occupied nearest neighbors and one vacant nearest 
neighbor are active  (filled circle) whereas   all other particles  are inactive (open circle). The active particles hop to the  {\it only} vacant nearest neighbor they have.}
\label{fig:model}
\end{center}
\end{figure}

The  two-chain   model ($M=2$) is defined on a periodic one-dimensional ladder of length $L$, 
i.e. total number of sites is $2L$ labelled  by $i=1,2,\dots, 2L.$ Each site  $i$ of  the ladder 
is  either   vacant  or be occupied by at most one particle;  correspondingly    the   site   $i$
is denoted by $s_i=0,1.$  A   generic configuration of the  system is   thus  represented by, 
$$C \equiv\left\{ \begin{array}{ccccc} .. & s_{i-1} & s_{i} & s_{i+1} & ..\cr .. 
& s_{i-1+L} & s_{i+L} & s_{i+1+L}  & .. \end{array}\right\}.$$
Particles   are allowed to  hop to the neighboring vacant site  with a rate 
that depends upon the total number of  occupied neighbors: those  having exactly two occupied 
neighbors (out of three) hop  with unit rate to the {\it only} vacant neighboring site they have.
A schematic description of the dynamics is shown in Fig. (\ref{fig:model}).  All the   possible  
hopping  scenario  are  listed below, where  active   particles are   shown  with a $\hat1$ 
and  sites  marked  $*$  can  be  in any state,  vacant  or occupied.  
\bea
\left\{ \begin{array}{c}*1*\\1\hat10\end{array}   \right\} \to 
\left\{ \begin{array}{c}*1*\\101\end{array}   \right\}; ~~
\left\{ \begin{array}{c}*1*\\0\hat11\end{array}   \right\}\to
\left\{ \begin{array}{c}*1*\\101\end{array}   \right\} \cr \n \\
\left\{ \begin{array}{c}1\hat10\\ *1* \end{array}   \right\} \to 
\left\{ \begin{array}{c}101\\ *1* \end{array}   \right\}; ~~
\left\{ \begin{array}{c}0\hat11\\ *1* \end{array}   \right\}\to
\left\{ \begin{array}{c}101\\ *1* \end{array}   \right\} \cr \n \\
\left\{ \begin{array}{c}1\hat11\\ *0* \end{array}   \right\} \to 
\left\{ \begin{array}{c}101\\ *1* \end{array}   \right\}; ~~
\left\{ \begin{array}{c}*0*\\ 1\hat11 \end{array}   \right\}\to
\left\{ \begin{array}{c}*1*\\ 101 \end{array}   \right\} 
\label{eq:ladder_dyn}
\eea
Note that, unlike CLG on a  ladder studied  in \cite{B16}, here  the dynamics is  restricted  
to follow  deterministic  hopping.   

It is evident from the dynamics that total number of particles $N$ as well as particle density $\rho=\frac{N}{2L}$ is conserved. It can be understood that number of active particles (i.e. which can hop) in the system   depends on density of particles  $\rho.$  For low densities  all the  particles will be able to organize themselves  such that none of them  are surrounded by two occupied  nearest neighbors. Hence activity in the system will cease and the system   is expected to  fall into an absorbing  state.  On the other hand for large
densities,  many particles   would have    more than one  occupied  neighbor and  
hence system   can  remain  active.   Thus   one expects an  absorbing state  phase transition (APT)   to take place  when density of the system  is decreased below a critical threshold  $\rho_c.$   Our aim is   to characterize  the critical  behaviour   of  this  APT. 

For CLG on a ladder with the  deterministric  dynamics  (\ref{eq:ladder_dyn}), when an active particle at  site $i$
hops to the vacant nearest neighbor it creates a vacancy  at $i$ which  is now  surrounded  by  occupied sites.
Thus  particle  hopping can  never create  additional consecutive $0$s (CZs), neither   in horizontal  nor 
in vertical directions.   The existing consecutive $0$s, if present in 
the initial  configuration, can only decrease with  time. Thus, starting from
any initial   configfuration the system   would   reach a  stationary state, with minimum number  of  CZs. Since, 
for density $\rho<\frac12$  all  configurations   must  have  some  CZs, the stationary state  is expected to be absorbing. On the other hand,   when density  $\rho\ge  \frac12,$  this  dynamics   (\ref{eq:ladder_dyn})   is  expected to   get rid of all CZs present  in initial     configuration   and  the stationary state, like   1DCLG \cite{UrnaPK}, would be  devoid of CZs. 
Thus  the  stationary  configurations of the system   are   composed of   rungs (the vertical  supports) which do  not have any CZs. Explicitly,   among    the  four  possile   rungs $\bfrac{0}{0}$, $\bfrac{0}{1}$, $\bfrac{1}{0}$ and $\bfrac{1}{1},$ the   stationary   configurations   of  the system    with   $\rho\ge \frac12$  are   composed 
of only three; the rung $\bfrac{0}{0}$  must be absent.  To  keep  track   of   the number of partricles, we denote the rungs by two indices $n$  and $k$,    $n$   the   number of particles in the rung, and $k=1,2,\dots, \kappa_n$ is a  running index   that distinguishes  different  rungs   in a   given $n$-particle sector. Here,   
$\kappa_1=2, \kappa_2=1$   denotes the  number of   rungs in  $n=1,2$ particle sectors respectively.

The configurations 
in the stationary state  are  now, 
\be
C\equiv   \{n_1k_1, n_2k_2, \dots  ,n_Lk_L\} \equiv     \{ n_ik_i\}. 
\ee
However, any  arbitrary  combination of these
three  rungs  are  not allowed   in the  stationary state  as they may  produce  CZs. For  example repetion of 
rungs  $\bfrac{0}{1}$ or  $\bfrac{1}{0},$  which   create    CZs   in horizontal direction, must   be   absent in the stationary state.

It is   evident that  in absence  of CZs, a particle   that hops from site $i$ to  $j$   would   find   that  two   of  its neighbors   at  $j$  are   already occupied   and  thus,     hopping of  this   active  particle  at site $j$ to the  vacant neighbor $i$ is also allowed by the  dynamics  (\ref{eq:ladder_dyn}). So, the stationary   dynamics   satisfy  detailed  balance   with  steady state  weights given by,  \be
 w(C)  = \left\{  \begin{array}{cc}  0 & {\rm if ~ CZs~ are ~ present}  \cr 1  & {\rm otherwise} \end{array}\right.
\label{eq:weight}
\ee
In the follwoing,  we  construct a transfer matrix  $T$  so that 
\bea
w(\{n_1k_1, n_2k_2, \dots  ,n_Lk_L\}) = \prod_{i=1}^{L} \la {n_i}_{k_i}|T | {n_{i+1}}_{k_{i+1}} \ra,
\label{eq:wCT}
\eea
where the ortho-normal basis vectors  for   the  transfer matrix   that  corresponds to different rungs are, 
\bea
\bfrac{1}{0}  \equiv |1_1\ra;  \bfrac{0}{1} \equiv |1_2\ra; ~ {\rm and}~   \bfrac{1}{1} \equiv   
|2_1\ra, \eea
Here, again we use  a  notation  $|n_k\ra,$   with  $n$ being the   number of particles in the rung and 
$k=1,2,\dots$ is a  running index   that distinguish   different  rungs   in a   given $n$-particle 
sector.  To ensure that  the  weight  of all  those configurations that produce CZs in  horizontal direction
are  zero,  we  must  set 
\be
\la 1_1| T| 1_1\ra=0 = \la 1_2| T| 1_2\ra.
\ee
Explicitly, the $3\times3$   transfer matrix   is given by, 
\be
T= \left( \ba{ccc} 1&1&1\cr 1&0&1\cr1&1&0 \ea \right).
\ee
In fact the weights, as writen in Eq. (\ref{eq:wCT}),  enesures  that the steady state  of the  CLG on a
ladder    has a  matrix product form where  each rung   is represented by a matrix, 
\bea
\bfrac{1}{0}  \equiv |1_1\ra \la 1_1|T;  
\bfrac{0}{1}   \equiv|1_2\ra \la 1_2|T;  \bfrac{1}{1} \equiv  |2_1\ra \la 2_1|T. 
\label{eq:rung_rep}
 \eea

\begin{figure}\vspace*{.2 cm}
\begin{center}
\includegraphics[width=8.5 cm]{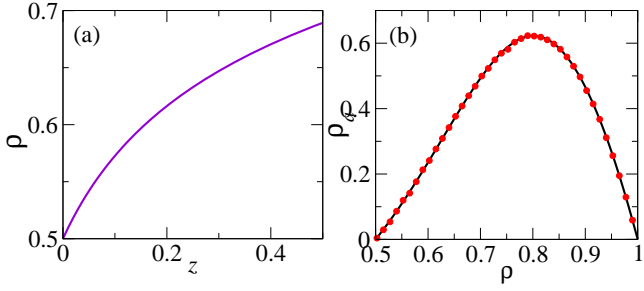}
\caption{(Color online)  (a) Plot  of   density $\rho$ as a function of $z:$  $\rho$ approaches the  critical value $\rho_c=\frac12$  as $z\to0.$     (b) Plot of order parameter $\rho_a$ (i.e. steady state density of active particles) versus $\rho.$  $\rho_{a}$ becomes nonzero above critical density $\rho_{c}=\frac{1}{2}$. 
For $\rho=1,$ all the sites are occupied  and  thus $\rho_{a}=0$.}
\label{fig:rhoa_M2}
\end{center}
\end{figure}

The steady state probability of any configuration 
\be
P_N(\{n_ik_i\}) =  \frac{w(\{n_ik_i\})}{Q_N}.  
\ee
Here, $Q_N$ is the canonical partition function,
\be
Q_N= \sum_{\{n_ik_i\}}  w(\{n_i k_i\} ) \delta(\sum_i n_i - N),\label{eq:M2_CE}
\ee
which, in  this   model,  counts  the number of  recurring configurations of a system  of size  
$2L$ containing $N$ particles. 
It  is convenient to work  in the grand canonical  ensemble (GCE) where  density of the system can be tuned by 
a  fugacity   $z.$ The   partition function in GCE is $Z(z)=   \sum _{N=0}^\infty z^N Q_N$; from Eqs. (\ref{eq:wCT}) and (\ref{eq:M2_CE}),
\bea
Z(z)&=& Tr[C(z)^L]; {\rm where}, \cr ~~  C(z) &=& \sum_{n=1}^2 z^n \sum_{k=1}^{\kappa_n} |n_k\ra \la n_k|T 
=  \left( \ba{ccc} z^2&z^2&z^2 \cr z&0&z\cr z&z&0  \ea \right). \label{eq:Cz_M2}
\eea
The  eigenvalues of $C(z)$ are,
\bea
\lambda\pm =\frac{z}{2}(1+z\pm\sqrt{z^2+6z+1}); \lambda=-z.
\label{eq:eigenvalues}
\eea
In the thermodynamic  limit $L\to \infty,$   the  partion   function   gets the dominant contribution from 
$\lambda_+,$ the largest eigen value  of $C(z),$
\be
Z(z) \simeq \lambda_+(z)^L.  
\label{eq:Z_M2} 
\ee
The average steady state density of the system   is then 
\be
\rho(z)  =\frac{z}{2} \frac{\partial}{\partial z} ln\lambda_+(z)\label{eq:rho_M2} 
\ee
In Fig \ref{fig:rhoa_M2}(a)  we   plot   $\rho$ as a function   of $z;$  it approaches  a  finite value $\rho_c=\frac12$ as  for  $z \to 0.$ Hence, the critical density below which the system goes to a absorbing state 
is  $\rho_c = \frac{1}{2}$. In this   critical limit, 
\be
\lim_{z\to 0} \rho_{z} \simeq\frac12+ z -4z^2+\c O (z^3) \Rightarrow  z \simeq  (\rho- \rho_c).
\label{eq:rho_expansion1}\ee

Above critical density $\rho>\frac12,$  the system remains in active phase. 
To measure activity,  the  density  of active particles  $\rho_a$,  as a 
function of tuning parameter $\rho,$ we calculate  the  probability  that 
an occupied site   is active   in  the steady  state.  To determine whether  an occupied  site  is  active, one must  check the occupancy status of all  its neighbors; thus  the activity  $\rho_a$ is  the steady state 
average  of the  following three-rung-local-confugurations, 
\bea
\rho_a &=&2\la\ba{l}110\\011\ea\ra + 2\la\ba{l}011\\110\ea\ra + \la\ba{l} 111\\101\ea\ra   + \la\ba{l}101\\111\ea\ra+ 
\cr &+& \la\ba{l}111\\110\ea\ra +\la\ba{l}110\\111\ea\ra +\la\ba{l}011\\111\ea\ra  + \la\ba{l}111\\011\ea\ra.
\label{eq:rhoaM2}
\eea
A  factor $2$ in first two terms   indicate that  these local configurations have two active sites.  
  Let us calculate  the   first term   explicitly, others   can 
be calculated in a similar way. 
\bea
\la\ba{l}110\\011\ea\ra &=& \frac{1}{Z(z)} Tr[  z |1_1\ra \la 1_1|T z^2 |2_1\ra \la 2_1|T z|1_2\ra \la 1_2|T 
C(z)^{L-3}] \cr
&=& \frac{z^3}{Z(z)} \la 1_2|C(z)^{L-2}|1_1\ra.\label{eq:rhoaM2-II}
\eea
It is  evident that the first  two terms   of (\ref{eq:rhoaM2})  gives rise to the lowest order 
terms in $z,$  as these  three-rung-configurations  have four particles  
in total,   whereas the others have  five  (each particle contribute a   factor $z$).

All  the  terms   of   (\ref{eq:rhoaM2}) can be calculated in a similar way, as in (\ref{eq:rhoaM2-II}).
The exact  expression of  $\rho_a $ as a function of $z$ is  long and we   do not  present  it here, but 
a  parametric plot  of $\rho_a(z)$ as  a function of $\rho(z)$ is  shown in Fig. \ref{fig:rhoa_M2}(b).  
It clearly shows that $\rho_a$ vanishes  linearly as  the density approaches the critical limit 
$\rho\to \rho_c= \frac12,$  i.e., $\rho_a \propto (\rho-\rho_c)$  and thus    the order parameter   exponent  of the absorbing phase transition is  $\beta =1.$   In the  same figure,  the  points   represent  the value of    $\rho_a$   obtained   from Monte-Carlo simulation of  the  restricted  CLG  on a  ladder,   for  a system  size $L=10^3.$

In fact, to obtain the order parameter exponent $\beta$, it is enough to calculate  one of the  first 
two terms  in the expression  of $\rho_a$ in Eq.   (\ref{eq:rhoaM2}),   which  are  lowest order in  $z,$ 
because   in the  critical limit  $z\to 0$  these  terms, if    turns  out to be  nonzero,   
contribute dominantly.  We  consider, 
\be
\rho_a^* \equiv \la\ba{l}110\\011\ea\ra = \frac{z^3}{Z(z)} \la 1_2|C(z)^{L-2}|1_1\ra.
\ee
To the lowest order (for system with even number of sites),  from Eq. (\ref{eq:Cz_M2}) we have  
\bea 
&&\la 1_2|C(z)^{L-2}|1_1\ra \cr
 &&~~~~~~=  z^{L-2} \la 1_2|T|1_1\ra \la 1_1|T|1_2\ra  \dots  \la 1_2|T|1_1\ra = z^{L-2}.\n
\eea
and $Z(z)  \sim z^L$   (from Eq. (\ref{eq:Z_M2})). Thus, 
$\rho_a^*  \simeq  z.$
Again,  from  Eq. (\ref{eq:rho_expansion1}),  $z \propto (\rho-\rho_c),$  implying    $\rho_a^* \propto (\rho-\rho_c)$ 
and   thus  $\beta=1.$ In Fig. \ref{fig:evnChain}  we have  shown  a plot of $\rho_a^*$ as a function of   $\rho,$   (solid line) along with the  same obtained from  Monte-Carlo simulations  of a  system of size $L=1000.$

In the critical limit the total activity $\rho_{a}\simeq z \simeq (\rho-\rho_c)$,  can also be obtained  
directly from the Taylor series expansion of $\rho_a$.  However,  the number of  active-three-rung configurations
that contribute to $\rho_a$  rapidly increase  for larger $M$-chains, and it is convenient to  calculate  $\beta$ from   $\rho_a^*,$ rather than from $\rho_a.$

Now  we turn our attention to  the  density correlation function.  It is evident  from  Eq. (\ref{eq:rung_rep}) that,  
the matrix representation for the  particle ``$1$''   and the vacancy ``$0$''  are respectively  $DT$ and $ET$ where
matrices $D$ and $E$ are given by,  
\bea
D=|1_1\ra \la 1_1| + |1_2\ra \la 1_2| +  |2_1\ra \la 2_1|; E =|1_1\ra \la 1_1|
\eea
Thus,  the average  density of the system is  $\rho(z) = \frac{Tr[DC(z)^L]}{Tr[C(z)^L]};$ it is straight forward 
to  show that  in the thermodynamic limit   this expression  is equivalent to Eq. (\ref{eq:rho_M2}).  The  density 
correlation function  is now, 
\be 
g(r)= \la s_i s_{i+r}\ra - \rho^2 =\frac{Tr[ DC(z)^rDC(z)^{L-r}]} {Tr[C(z)^L]}. 
\ee
In the thermodynamic  limit 
\bea 
g(r) \propto \left(\frac{\lambda_-}{\lambda_+}\right)^r  = e^{-\frac r\xi}; \xi^{-1} = 
|\ln\frac{\lambda_+}{\lambda_-}|.
\eea
From Eq. (\ref{eq:eigenvalues}),  it is evident that  the correlation length $\xi$ diverges  in the 
critical limit $z\to 0,$
\be
\xi  \simeq \frac1z  = (\rho-\rho_c)^{-\nu}; ~{\rm with~exponent~} \nu =1.
\ee
Any  rung-rung correlation function, or   the  correlation functions   for activity  also   decay  
exponentially (not shown here) with the same length scale $\xi.$   Since  {\it at}  the critical point one expects  power 
law correlation, $g(r) \sim  r^{2-D-\eta}$   we  conclude that  for this quasi-1D system   $\eta=1.$

The critical exponents  that we obtained for the  restricted CLG  model on a ladder is thus characterized by 
the critical exponent $\beta=1=\nu=\eta,$  which  is same  as the  CLG model in 1D.  Previous studies  of CLG 
models on the  ladder \cite{B16}  exhibit absorbing transition in  DP  universality class due to  the fact that 
the  dynamics  of that  model  was   essentially  stochastic,  in the sense  that  the  active  particles there 
may have  more than  one vacant neighbors,  and then    it must  choose one  of  them  randomly as  
the target site, and hop here.   Once the stochastic particle transfer is ceased, in the  present model, 
the critical   behaviour of the   absorbing  transition   becomes  same as that  of  1DCLG.

In the following   section we discuss  multi-chain system  and calculate  the   critical   exponents
of the absorbing transitions there.   We see  that the odd  and even number of chains  exhibit 
different  universal feature.

\section{Multi-chain system}
\label{sec:Multichain}
The multi-chain  models   are  straightforward  generalization of the  restricted CLG  on ladder  discussed 
in the previous section, but  their  critical behavior depends on $M,$ the   number of chains.  Formally  
we  start with a $M\times L$ square lattice  where each site  $i=1,2,\dots ML$ is  either vacant  ($s_i=0$) 
or   occupied by   one particle $(s_i=1).$  Further,  we assume periodic  boundary condition in both  $x$- and $y$-direction.
The  dynamics of the system   for $M>2$  is    similar to  the one defined on  a  ladder ($M=2$):  sites which have 
{\it exactly} one    
vacant  neighbor (i.e.,  three  other  neighbors  are occupied)   can  hop to the vacant site with unit rate. 
The rightward hop  of an active particle   is then
\be
\left\{ \begin{array}{c}\dots\\ *11\\1\hat 10\\ *11\\\dots\end{array}   \right\} \rightarrow
\left\{ \begin{array}{c}\dots\\ *11\\101\\ *11\\\dots\end{array}   \right\},
\ee
where *   represents an arbitrary occupancy  -vacant or  occupied-  and the  active particle is marked with a hat. 
Similary,   the active particle can also hop to  the left  or  upwards   or downwards  when  these  sites  are 
only   vacant  neighbor  of a  particle. 

This  dynamics conserves  the  total number  of  particle  $N$  or density $\rho=\frac{N}{ML}$ 
and,   like the  dynamics on  a ladder,  can not create  consecutive  $0$s but  destroy the ones   
present in the system.  Thus one expects that,  consecutive $0$s are absent in the  steady state \cite{footnote1}.
Thus we  work in  $\rho> \frac 12$   regime   and assume to start with a 
initial  configuration  which  does not have any  consecutive 0s. 
Since  the system, once  transit  from  a  configuration with higher number of CZs,   will   never comes  back to 
visit it again (as    generalization   of  CZs   are  not allowed  by the dynamics)  
it would be economic  in terms of simulation time  to start with an initial configuration
which  does not have any  consecutive 0s.   For these models  we  call such   ICs  as  natural initial
configurations (natural ICs) and  it is  certainly possible to  create   such  configurations
for $\rho\ge  \frac 12;$ we choose to discuss   $\rho>\frac12$  case in   more details  and show that
the  critical density is  $\rho_c\ge \frac12$  for all $M.$   These  models for  $M>2$ have
some subtle  features for  $\rho>\frac 12$  which were not present  in  $M=1$  \cite{UrnaPK}  
or $M=2$ (previous section); we will  discuss  these issues  in section \ref{sec:conclusion}  in some details.

It is easy to see that  in absence of CZs,  if  the dynamics allows a transition from   any 
configuration   $C$ to another one $C'$   it also   allows  the reverse transition   $C'\to C.$  Since   any such  transition occurs with  unit rate, the steady sate  must  satisfy detailed  balance, with  steady   state  weight   $w(C) =1$  for 
all $C$   which are  devoid of   CZs. Thus,   all  the  configurations (devoid  of CZs) in the supercritical regime are  equally likely.  Our  first step  is  to enumerate    such  configurations. 

Any $M$-chain system  of size $L$ consists of $L$ rungs, which are  the vertical  supports.  Since we  want to 
construct configurations  which are devoid of CZs,   we  must primarily  ensure that  every rung  must not contain  any CZs. Let $d_M$  be the 
the number   such rungs; clearly,   $d_M$ is  same as  the  number of  allowed  configurations in  the    steady state of 1DCLG model \cite{UrnaPK}  on system size $M$  with   $\frac M2$ or more particles.  This   is because, for  density   larger than  $\frac12$ the 1DCLG   model  lead   to a  steady  state  where there  CZs are  absent. 
The steady state  weights of these  models  can be expressed in  a  matrix product form \cite{UrnaPK};  the grand canonical partition function with a fugacity  $z$ that    controls the particle density of  the  $1D$  chain for a 
system size $M$ is  given by  
\be Z_{1D}(z) =  Tr [ \left(\begin{matrix} z  & 1  \\  z & 0 \end{matrix}\right)^M].
\ee
For $z=1,$  the partition   function  counts   all   possible  configurations  of the system  irrespective of its  
density. Thus, 
\bea
d_M &=& Z_{1D}(1) =Tr [ \left(\begin{matrix} 1  & 1  \\  1 & 0 \end{matrix}\right)^M] \label{eq:1110}\\ 
  &=& \frac{1}{2^M} [ (\sqrt 5  +1)^M  + (\sqrt 5  -1)^M]. 
\eea
In fact,   the matrix  that appears   in Eq. (\ref{eq:1110})    is simply the   transfer matrix  which is used to construct a 
binary  string  which does not posses CZs.    Also note    that  the asymptotic form of $d_M$  is 
\be
d_M   \simeq  \phi^M ~~  {\rm where} ~  \phi= \frac{ \sqrt 5  +1}{2} 
\ee
is the golden ratio. 

 The  $M$-chain system is   composed of  $d_M$ different    kind of rungs, but any  arbitrary arrangement of  rungs  is not allowed in the steady state. This  because, the  rungs  themselves   does not contain any CZs, but 
 any  arbitrary  placement   of rungs  could   generate   CZs on  horizontal bonds.  Our aim  would be to construct a
 transfer matrix considering each of the rungs as  basis vectors, which   would  automatically take care of the forbidden  arrangements.  Let   us  categorize  the collection of $d_M$  rungs   with respect to the number of  particles  they have; in  the $n$ particle sector we have  say  $\kappa_n$    rungs   labeled   by $k=1,2,\dots \kappa_n.$ 
 Thus, \be  \sum_{n=\nu}^M  \kappa_n =d_M, \ee  where $\nu$  is the minimum number particles  in a   rung.   
 Since  the  rungs  do  not contain    CZs  in vertical direction,   the minimum number particles  
 in rung  is
 \be
 \nu  = \lfloor\frac{M+1}{2}\rfloor=\left\{  \begin{array}{cc}   M/2  &  for ~M=even  \\ (M+1)/2  &  for ~M=odd \end{array}   \right., 
 \ee
and the maximum    number  is $M.$  The exact   value of $\kappa_n$ (number of rungs that contain exactly $n$ particles and of course $M-n$ vacant sites)  is the coefficient of $z^n$ in  the  Tayler's series expansion of   $Z_{1D}(z)$ about $z=0,$ 
\be
\kappa_n =  \{z^n\}  Z_{1D}(z).
\ee
For  some $n$ it is straightforward  to  calculate  $\kappa_n.$  For example, for $n=M$  we have 
$\kappa_M=1$ (the rung is filled with 1s), for  $n=\nu,$  $\kappa_\nu =2$  when $M$ is odd 
(alternative  sites  are  occupied,  starting with  0 or 1) and    $\kappa_\nu =M$  for odd $M$   
(with $n=\frac{M+1}{2}$  particles, one of the $M$ vertical   bonds  of the   rung   must  
have consecutive 1s).

At this  stage we use a systematic ordering of the rungs,   which  can act as  the basis-vectors for the transfer matrix. We represent the rungs  by $\{n_k\}$ where   $n,k$ are   integers - $n$   varies   in the range 
$(\nu,M),$ and $k,$ for a  given $n,$ varies    in the range  $(1,\kappa_n).$ The  standard basis for  the 
transfer matrix  is  a set of orthonormal vectors 
\bea
\{ |n_k\ra\} \equiv && \{|\nu_1\ra, |\nu_2\ra, \dots,|\nu_{\kappa_\nu}\ra \cr&& \dots\cr&& 
|n_1\ra, |n_2\ra,  \dots  |n_{\kappa_n}\ra, \dots \cr&&
\dots     \cr&&
|M_1\ra \}.
\eea   
In this  basis,  the  elements of the transfer matrix  are  nonzero, $\la n_k| T |n'_{k'}\ra = 1,$ when 
two rungs   $|n_k\ra$  and  $|n'_{k'}\ra$  as neighbors  do not  produce any  CZs 
in  the  horizontal  direction,  i.e.  if  the  one of   rung has  $0$s at certain  positions, the  other  
must  have $1$s at that position. 
\be
\la n_k| T |n'_{k'}\ra = \left\{  \begin{array}{cc}  1  &  {\rm if}  |n_k\ra,|n'_{k'}\ra {\rm do~ not~ generate ~ CZs}  \\ 
0  & {\rm otherwise} \end{array}   \right..\n
\ee
It is easy to  obtain the   transfer matrix manually  for small $M,$  but the dimension of the  matrix $d_M\sim \phi^M$ grows  
exponentially  and  quickly the calculation becomes tedious.  However it can be  computed   numerically  noticing  the fact 
that for any  two  $M$-bit binary strings $s$ and $s'$ which does not  have consecutive  zeros,  the operation $\tilde  s \& \tilde {s'},$ where $\&$ and \~~ represent   bit-wise AND  and NOT  operations  respectively,   gives a   nonzero  value  only when there is at least   one  spatial position where both strings have a $0.$  

It is  easy to see   that   $T^L$ generates   all  possible  configurations devoid of  CZs, irrespective 
of the  number of particles ($1$s). To   describe   $M\times L$  system  with a  conserved   particle number $N$ (or conserved 
density  $\rho = \frac{N}{ML}$) we   introduce a  fugacity  $z$  and  write  the partition function  in grand   canonical ensemble as  
\be
Z(z) = Tr [C(z) ^L];  ~  \la n_k| C(z) |n'_{k'}\ra  =  z^n \la n_k| T |n'_{k'}\ra..
\label{eq:GCE1}
\ee
Since the  minimum  number of particles  in   any of the rung  is  $\nu,$ we can   expand 
$C(z)$ as  follows, 
\be
C(z) =  \sum_{n=\nu}^{M}  z^n C_n  \label{eq:Cn}
\ee
where  matrices  $C_n$  are independent of $z.$ 
The description of  grand canonical ensemble is  incomplete,  unless we specify the  density  as a   function of 
fugacity $z.$   Density  of the   $M\times L$ system can be calculated by  taking  trace ($Tr[.]$)   over 
all  configurations where one specified  site  of the   system   is occupied.  
Since the  rung  ${n_i}_{k_i}$   at site  $i$  is only a binary string   $\{s_i, s_{i+L}, \dots s_{i+(M-1)L}\}$  with  $\sum_{j=0}^{M-1} s_{i+jL} =n_i,$ we can associate an unique decimal value  ${\cal D} (n_k) = \sum_{j=0}^{M-1}  2^{j} s_{i+jM}$ to it; the decimal value is an odd integer   if   first site of the rung is  occupied. 
%
%
Thus   by defining a  diagonal matrix, 
\be
D= \sum_{n= \nu}^M  \sum_{k=1}^{\kappa_n}  |n_k \ra \la n_k |  \delta \left( 1- {\cal D} (n_k) ~{\rm mod}~ 2\right) 
\label{eq:Dmatrix}
\ee
we   get  the density of the system as, 
\be 
\rho(z) = \frac1{Z(z)}Tr[D C(z)^L] . \label{eq:rhoD}
\ee
Of course,  one standard way  one  calculate  the density is as follows. If  the  largest  eigenvalue of $C(z)$ is 
$\lambda(z),$  in the thermodynamic limit   $Z(z) \simeq \lambda(z)^L$  and    density     is 
\be 
\rho(z) = \frac{z}{M} \frac{d}{dz} \ln \lambda(z). \label{eq:rho_diff}
\ee
However, when the dimension of the transfer matrix is large (which is indeed the  fact  as   the dimension   
$d_M \sim \phi^M$)  it is advantageous to calculate   $\rho(z)$   numerically, using   
Eq. (\ref{eq:rhoD}).

In  the following we see  that  the critical density $\rho_c$ where 
the $M\times L$ system    undergoes  a  non-equilibrium  phase transition   from an active  to an absorbing state  is 
\be 
\rho_c =  \lim_{z\to 0} \rho(z)  
\ee
and  the critical behaviour of the system depends on  how  the  partition function  and other observables 
depend  in the $z \to 0$ limit; in this regime   contribution  from  matrices $C_\nu$  and  $C_{\nu+1}$ are most important. 

\subsection{Steady state    in matrix product form}
The   steady state   average  of   different observables,  can be calculated  easily, if we  write  the steady state weights  of 
the configurations   in  a  matrix product form.  
Every configuration of the system   is composed  of $L$ rungs.  Denoting   a  rung $n_k$  by 
by a  matrix $R(n_k)$ (in  total  there  are  $d_M$  number of different matrices)  
the steady state  probability of a  configuration  $\{n_ik_i\}$   can be   written  in a 
matrix product form using a matrix product ansatz,  
\bea
&&P(\{ {n_1k_1}, {n_2k_2}  \dots   {n_Lk_L} \} \cr
&&= \frac{1}{Q_N} Tr[ \prod_{i=1}^L R({n_i}_{k_i})] \delta \left(\sum_{i=1}^L n_i-N \right)\nonumber
\eea
where  the $\delta$-function 
ensures  conservation of  the number of particles  $N,$  and  $Q_N$ is the canonical  partition function, 
\be
Q_N =  \sum_{\{n_i=\nu\}}^M \sum_{\{k_i=1\}}^{\kappa_{n_i}}\ Tr\left[ \prod_{i=1}^L R({n_i}_{k_i})\right] \delta \left(\sum_{i=1}^L n_i-N \right).\n
\ee
The grand canonical    partition function is then, 
\bea
Z(z) = \sum_{N=0}^\infty  z^N Q_N  =  Tr\left[ \left(\sum_{n=\nu}^M \sum_{k=1}^{\kappa_{n}}  z^n R(n_k) \right)^L\right]\nonumber
\eea
Comparing this  with Eq.  (\ref{eq:GCE1}), we get matrices, 
\bea
C(z) =\sum_{n=\nu}^M z^n \sum_{k=1}^{\kappa_{n}}  R(n_k) \label{eq:MPA2} \\
~{\rm and} ~ R(n_k) =  |n_k\ra \la n_k| T   \label{eq:Representation}.
\eea
Equation  (\ref{eq:Representation}) is very important   to us,  as   any  explicit matrix 
representation    is useful  for the calculation of  observables.  For example,  the steady state 
average    of a  particular   rung  $\bar n _{\bar k}$   is 
\be 
\la \bar n _{\bar k} \ra =  \frac{Tr[ z^ {\bar n} R(\bar n_{\bar k})  C(z)^{L-1}]}{Tr[C(z)^L]} 
=  \frac{\la\bar n\bar k| C(z)^L|\bar n\bar k\ra   }{Tr[C(z)^L]}.
\ee
A comparision of Eqs. (\ref{eq:MPA2}) and   (\ref{eq:Cn})  gives, 
\be
C_n = \left[\sum_{k=1}^{\kappa_n}  |n_k\ra \la n_k| \right] T  = \Pi_n T, \label{eq:CnRep}
\ee
where  $\Pi_n$  is the   projection operator, defined by the  term  within the bracket $[.],$ which 
projects out  all the rungs   having  exactly $n$  particles. 

One  important  observable we would be interested in  is the  order parameter of the absorbing phase transition,  
namely  activity. Writing   a matrix  representation   for it is not that  simple,  as  constructing all possible arrangements  of
 the rungs that  can create  active sites  is  not   possible for general  $M$; for $M=2$, as  we have discussed in  the previous 
section,  there  are  eight    3-rung configurations  which  have at least one active site.  However  one can infer  
about the behaviour  of  the order parameter   at the  critical point  easily  by considering only  any of the 
3-rung-configuration which  has the  minimum number of particles  which   contribute   to the  lowest order in $z.$ However,  we have already mentioned, 
the critical behaviour of  the system  with   odd  number of chains    are different  from 
the  same   with even $M$;  we  discuss these  two cases  separately  in the  following  two subsections. \\

\subsection{Restricted CLG  on odd number of chains}\label{sec:odd_chain}

 For odd  number of chains,  $M= 2m +1,$   the minimum number  of  particles on a  rung  (which does not have CZ's) 
 is   $\nu =  m+1.$ There  are  exactly $M$-rungs which has $(m+1)$   particles(1s)  and  $m$   holes (0s), thus each one   contain  exactly one  consecutive 1s in the  vertical direction. We  denote  these rungs  as 
\be
|\nu_1\ra= \left(\bm1\\1\\0\\1\\0\\1\\0\\ \vdots\em\right),
|\nu_2\ra= \left(\bm0\\1\\1\\0\\1\\0\\1\\ \vdots\em\right),
\dots,
|\nu_{k_\nu}\ra= \left(\bm1\\0\\1\\0\\1\\0 \\\vdots \\1\em\right).\label{eq:odd_rungs}
\ee  
Our first aim is calculate  the critical density $\rho_c$ for   CLG  dynamics  on a system with 
odd   number of chains.  In fact, since one  can construct configurations   of 
 $L$-rungs  (without    any  CZs)   using only the rungs containing  $\nu$  particles (like $\{\nu_1, \nu_2,\nu_1,\nu_2\dots\}$)   the steady  state  density  of the system  can    not   decrease below $\nu/M,$  and  one  expects   the  critical density  to be $\rho_c \ge \frac{m+1}{2m +1}.$  We show below that  
$\rho_c = \lim_{z\to 0} \rho(z)  = \frac{m+1}{2m+1}.$  

In the $z\to 0$ limit,  the partition function is, 
 \bea
Z(z)&&=z^{\nu L}  Tr[  (C_\nu +  z C_{\nu+1}  + {\cal O} (z^2) )^L ]\cr
  &&= z^{\nu L}\left( Tr[C_\nu^L] + z \sum_{k=0}^{L-1} Tr[ C_\nu^k C_{\nu+1} C_\nu^{L-1-k}] + {\cal O} (z^2) \right) \n
 \eea
 Thus the critical density is, 
 \be  
 \rho_c = \lim_{z\to 0} \rho(z)=  \frac{Tr[D C_\nu^L]}{Tr[C_\nu^L]}
 \label{eq:rhoodd}
 \ee
 Now,   since  the   $M$ rungs  (vectors)  in $(m+1)$-particle sector  are related  to each  other  by a rotation symmetry
 (with  respect to the  position  of a  {\it single}  consecutive 1s  in the vertical direction),  
 $\la \nu_k| C_\nu |\nu_k\ra =\la \nu_1| C_\nu |\nu_1\ra$  for any  $k=1,2,\dots \kappa_\nu=M.$  Thus  
 \be Tr[C_\nu^L] = \kappa_\nu \la \nu_1| C_\nu^L |\nu_1\ra. \label{eq:rho1}
 \ee   Again matrix 
 $D,$ defined in  Eq. (\ref{eq:Dmatrix}),   projects  out only  those rungs which has $1$ in the  first  position irrespective of 
 the  total number of particles.  Thus, 
 \be
 Tr[DC_\nu^L] =  \sum_{k}^\prime \la \nu_k| C_\nu^L|\nu_k\ra = \nu_{odd}\la \nu_1| C_\nu^L |\nu_1\ra,  \label{eq:rho2}
 \ee
 where  $\prime$  indicates  that  the sum   is restricted   to  consider only   those $k$ for which ${\cal D}(\nu_k)$ is an 
 odd  integer. The number of such rungs  in $(m+1)$-particle sector  is $\nu_{odd} =  m+1.$ 
 Thus   the critical density,   Eqs. (\ref{eq:rho1})   and (\ref{eq:rho2}),  is 
\be
 \rho_c  = \frac{m+1}{2m +1} =\frac{1}{M} \lfloor \frac{M+1}2\rfloor. \label{eq:rhoc_odd}
 \ee
   Further,    in $z\to 0$ limit,  using Eqs.  (\ref{eq:Cn}) (\ref{eq:rhoD}) we get   
 \bea
\rho(z)  &\simeq& \frac{Tr[D(C_\nu  + z C_{\nu+1})^L]}{Tr[(C_\nu  + z C_{\nu+1})^L]} \cr
&\simeq& \rho_c +  \gamma z   + {\cal O} (z^2),
\eea
where  $\gamma$ 
is a   constant independent of $z.$ Thus,  in this critical  limit,   \be  z  \propto  (\rho -\rho_c). \label{eq:zrho}\ee

 \begin{figure}
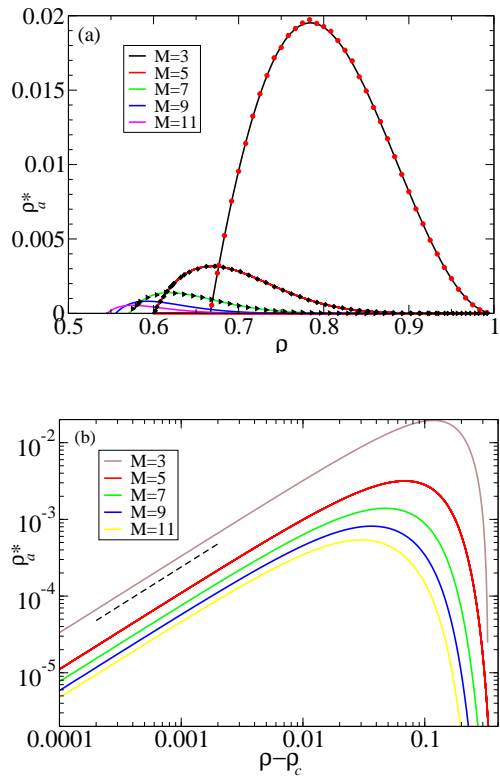
\vspace*{.2 cm}
\begin{center}
\includegraphics[width=6.5 cm]{oddChain.eps}\\ \vspace*{.78 cm}
\includegraphics[width=6.5 cm]{oddC_beta.eps}
\caption{(Color online)  The order parameter $\rho_a$ for is a  sum of  the steady state   average of several
3-rung-configurations, of  which   one  of the term,  which is lowest  order   in $z,$  is $\rho_a^*$
given by Eq. (\ref{eq:3rung_odd}). (a)  A parametric plot of $\rho_a^*(z)$  as a function of $\rho(z),$ 
calculated  following    the transfer matrix method (solid line)  is   compared with  the same obtained for  different  densities (symbols)  using  Monte-Carlo simulations  of  the restricted CLG dynamics (density conserving)  on a  $M\times L$ system, with  $L=1000$ and 
$M=3,5,7,9,11.$ Clearly,    $\rho_a^*$   vanish   at $\rho_c =  \frac{1}{M}\lfloor \frac{M+1}2\rfloor.$ 
(b) Log scale  plot of   $\rho_a^*$ as a function  of   $(\rho-\rho_c),$  along with a line  with unit slope (dashed line)      indicates  that $\rho_a^* \sim (\rho-\rho_c)^\beta $  with $\beta=1.$  
}
\label{fig:oddChain}
\end{center}
\end{figure}

We now proceed  to calculate   the order parameter $\rho_a(z),$   namely  the density of activity.   To  know that  a particle   at a  given site  is active, one need  to check that  all except one  of its  neighbor  is occupied.  For $M>2$,
since  every site   has four  nearest neighbors,  the  active particle  must have   three   occupied neighbors and one 
vacant neighbor; thus, one must consider {\it three}  consecutive rungs  to verify the occupancy of neighbors.  One  can place three different rungs several possible ways to construct active configurations (having at least one   active particle) which are devoid of  CZs; we  will not enumerate all these  configurations. 
To  know the behaviour of  activity $\rho_a(z)$  in the critical limit $z\to0,$ we need to   consider  only 
one of  active three-rung-configuration   with   minimum number of particles,  because  these configurations, being lowest order in $z,$  contribute dominantly   as  $z \to 0.$  In other words,  if $\rho_a(z)  = \sum_j  c_j z^{\alpha_j}$ with 
$\alpha_1<\alpha_2<\dots,$  an active-three-rung-configuration  leads  to  the  dominant  contribution  at $z=0,$ 
\be
\rho_a(z)  \propto  z^{\alpha_1}. \label{eq:rhoaz}
\ee
Comparing   Eqs. (\ref{eq:zrho}) and (\ref{eq:rhoaz}) we  obtain the order parameter exponent $\beta$, 
\be
\rho_a   \propto  (\rho-\rho_c)^\beta, ~{\rm where}~ \beta  =  \alpha_1.
\ee
For  odd $M$-chain,  the minimum number of particles  in  three  rungs   is  $3\nu =  3(m+1),$  i.e. when  ecah rung  has 
the minimum  $\nu$ number of particles; however   one can not   create an active configuration  only  with  these rungs. 
We  show that  an active configuration can be obtained with one extra  particle, i.e.  when one  of the three   rungs contain 
$\nu+1$ particle.  There  are many   such active configurations    with  $3\nu+1 = 3m +4$ particles; 
a  systematic   construction  for generic $M$  follows. This  construction is not unique, 
but a proof that  the steady state   average   of any such configuration  is   non-zero and 
it varies  as $z^{\alpha_1}$  in $z\to 0$ limit is enough   
for the   determine the critical   exponent  $\beta.$  

Let us take the  $|\nu_2\ra$ rung from Eq. (\ref{eq:odd_rungs})   and  put an  extra  particle on  the first    vacant site  on this  rung; this 
new  rung   belongs to  $(\nu+1)$ particle sector and we denote it as    $|(\nu+1)_1\ra.$  
Let  us take  the   active-three-rung-configurations   as  
\bea
\{\nu_1,(\nu+1)_1, \nu_3\} = \left( 
\bm1\\1\\0\\1\\0\\1\\0\\ \vdots\em
\bm1\\1\\1\\0\\1\\0\\1\\ \vdots\em
\bm1\\0\\1\\1\\0\\1\\0\\ \vdots\em
\right)
\label{eq:3rung_odd}
\eea
The  steady  state   average    of this  configuration for a given $M=2m+1$  is    
\bea
\rho_a^*(z) &=& \la\{\nu_1,(\nu+1)_1, \nu_3\} \ra =\frac{1}{Z(z)} \la \nu_1|C(z) |(\nu+1)_1\ra \cr
&\times& \la(\nu+1)_1| C(z) |\nu_3\ra\la \nu_3| C(z)^{L-2} |\nu_1\ra
\eea
In the $z\to 0$ limit,
\be
\rho_a^*= \frac{z^\nu z^{\nu+1} z^{\nu(L-2)}  \la \nu_3| C_\nu^{L-2} |\nu_1\ra }{z^{\nu L}Tr[C_\nu^{L}]} = A z
\ee
where  $A$  is a  positive  constant. This is because,  $C_\nu$ is a  positive symmetric matrix and 
$\la \nu_3| C_\nu^{L-2} |\nu_1\ra \ge  \la \nu_3| C_\nu|\nu_1\ra^{L-2}=1.$   Thus, for any odd $M=2m+1$  chain,
the order parameter $\rho_a,$ like $\rho_a^*$,    approach   to $0$ continuously as 
\be
\rho_a \sim (\rho-\rho_c)^{\beta} ; ~  \beta =1.
\ee
In Fig. \ref{fig:oddChain}(a) we   have shown   a parametric plot of  $\rho_a^*(z)$ as  a function of $\rho_a(z)$  
for different $M=3,5,7,9,11.$   The data points   in the same plot shows   $\rho_a^*$  obtained from  Monte-Carlo simulation of   the  restricted CLG dynamics  on    $M=3,5,7,9,11$ systems for  different  densities.  
The simulation  was done on a  system  of size $L=1000$ and    starting   from a  natural initial configuration.  
Clearly, the critical   density   for $M=2m+1$ is $\rho_c=  \frac{m+1}{2m +1}$  and it approach to $\frac12$ as 
$M$ increases. In \ref{fig:oddChain} (b) we  plot  $\rho_a^*$   as a function of  $(\rho-\rho_c)$   in 
log scale  to obtain  the  order parameter exponent $\beta=1$

 \subsection{Restricted CLG  on even  number of chains}\label{sec:even_chain}
 A special case of  even  $M=2m$ chain is the  ladder ($M=2$)  which is  discussed  in section \ref{sec:IIA}. 
 There,  we have explicitly calculated   the density $\rho(z)$ and the activity $\rho_a(z)$ and  found the order 
 parameter exponent $\beta = 1.$  Given, that   any odd  $M$ chain   undergoes an absorbing transition   with  exponent 
 $\beta=1,$  one naturally expects  that   the same must be true for  all  even  $M;$ this is, however, is not true.  
 Note that, a ladder   is a    very special case where    open  and  periodic  boundary conditions in  vertical  direction  results in same  lattice  structure. Further,  unlike  any $M>2$ system where every  site has four nearest neighbors,  the ladder has only three.  We  will see below that  $M=4$ is  also a  special case and it results  in $\beta=2,$  whereas  any even chain  with $M>4$ results  in a absorbing transition  with  exponent 
 $\beta  =3.$
 
  \begin{figure}\vspace*{.2 cm}
\begin{center}
\includegraphics[width=8.5 cm]{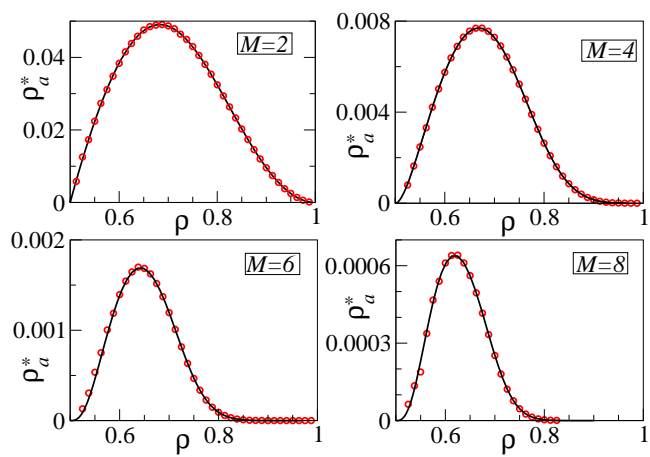}
\caption{(Color online) Parametric plot of $\rho_a^*(z),$  the    steady state  average   of  the  active  3-rung configuration  in  (\ref{eq:3rung_evn}) as a  function of $\rho(z)$  for  even $M=2,4,6,8$ along with  $\rho_a^*=\la\ba{l}110\\011\ea\ra$  for $M=2.$  Solid lines  are  obtained from  the  transfer matrix formulation
and the   symbols corresponds to the same obtained from  Monte-Carlo simulation of the  restricted CLG  on  $M\times L$ system with $L=1000.$  Clearly, $\rho_a^*$ vanishes at   $\rho_c=\frac12$ for all $M.$ 
}
\label{fig:evnChain}
\end{center}
\end{figure}

 For the even  $M=2m$, the minimum number of particles  in  the   rungs that   does not contain   consecutive 0s is  $\nu =m.$ There  are exactly  two rungs which  has  $\nu$ particles, i.e. $\kappa_\nu=2,$
 \be
|\nu_1\ra= \left(\bm1\\0\\1\\0\\1\\0\\ \vdots\em\right),
|\nu_2\ra= \left(\bm0\\1\\0\\1\\0\\1\\ \vdots\em\right). \label{eq:TWO}
\ee 
We  use   Eqs. (\ref{eq:rhoodd}), (\ref{eq:rho1})   and (\ref{eq:rho2}),   which  also  holds  true  when   $M$ is    even (can be checked easily) to calculate the critical density, 
\be 
\rho_c = \lim_{z\to 0} \rho(z)=  \frac{Tr[D C_\nu^L]}{Tr[C_\nu^L]}= \frac{\nu_{odd}}{\kappa_\nu}=\frac12.
\label{eq:rho_c_evn}
\ee
In fact, since the  rungs  are   devoid of CZs, the minimum density of a  configuration is  $\rho=\frac12$,  
obtained from, say $\{\nu_1,\nu_2, \nu_1,\nu_2...\}$  and one expects the critical denisty $\rho_c\ge\frac12.$  However, 
all  configurations  for density  $\rho>\frac12,$  are not active and  one need to check explicitly that  the 
minimum density is   the critical density.

Next we  focus on   the order parameter  $\rho_a.$  Here too, we need to   know  three  consecutive  rungs to 
identify whether  a particle  at a given site  is  active, i.e. the   active  site   must have  three occupied and 
one vacant neighbor. Of all such  three-rung configurations,  what contributes near the  critical point is  the 
active-three-rung-configuration that has   minimum number   of  $1$s.  Unlike odd $M$,  one can not   create an 
active-three-rung  configuration with  $3\nu +1$  particles, we need  at least  $3\nu +2$, i.e.  we need two  
rungs  with $\nu+1$ particles.  We  start with $M=4$  which is the  first  even chain system where   the  
lattice sites have   four  nearest neighbors and the extend it to  $M=6,8,\dots.$
Let us take  the  $|\nu_1\ra =(1,0,1,0,1,0,\dots),$  put a  particle at the first   
vacant site and move the particle at from 3rd to 4th  position and denote  this  rung as 
$|(\nu+1)_1\ra=   (1,1,0,1,1,0,1,0,\dots).$  Let  us  put a  particle  at the 2nd  vacant site of  $|\nu_2\ra=(0,1,0,1,\dots)$
and  denote it as $|(\nu+1)_2\ra=   (0,1,1,1,0,1,0,1,\dots).$   
The active  three-rung configurations for even  $M$   are  now  $\{(\nu+1)_1, (\nu+1)_2, \nu_1\},$

\be 
 M=4:\left( 
\bm1\\1\\0\\1\em
\bm0\\1\\1\\1\em
\bm1\\0\\1\\0\em
\right);~~
M>4 :
\left( 
\bm1\\1\\0\\1\\1\\0\\1\\0 \\ \vdots \em
\bm0\\1\\1\\1\\0\\1\\0\\1\\ \vdots\em
\bm1\\0\\1\\0\\ 1\\0\\1\\0 \\ \vdots\em
\right)
\label{eq:3rung_evn}
\ee
The  steady  state   average    of this  configuration for a given $M=2m$  is    
\bea
\rho_a^*(z) &=& \la\{(\nu+1)_1,(\nu+1)_2, \nu 1\} \ra \cr 
&=&\frac{1}{Z(z)} \la (\nu+1)_1|C(z) |(\nu+1)_2\ra   \la(\nu+1)_2| C(z) |\nu_1\ra \cr
&\times&\la \nu_1| C(z)^{L-2} |(\nu+1)_1\ra\cr
&=&  z^{2\nu+2}\frac{\la \nu_1| C(z)^{L-2} |(\nu+1)_1\ra }{ Tr[C(z)^{L}] } \label{eq:rhoa_evn}
\eea
If $|\psi\ra$, $\la \psi|$ are  respectively   the  right and  left normalized eigenvector of $C(z)$  corresponding   to the  largest  eigenvalue  
$\lambda_{max}= z^\nu \lambda(z)$,  in the thermodynamic 
limit  $L\to \infty$ one can write $\rho_a^*$ as 
\be
\rho_a^*(z) = \frac{z^2}{\lambda_{max}(z)^2} \la \nu_1|\psi\ra\la\psi|(\nu+1)_1\ra. \label{eq:master_rhoa}
\ee
This expression, being  independent  of system size $L$,  is  very useful in evaluating  
$\rho_a^*(z).$   The results for   different $M$  are shown in   Fig.   \ref{fig:evnChain}.
For $M=4,$    we  have  $d_M=7$ dimensional   matrix   $C(z) =  z^2 C_2 +  z^3 C_3  + z^4 C_4.$
Since  we are interested in the    $z\to 0$ limit it is sufficient to  take an approximation  
$C(z)  \simeq  z^2 [C_2 +  z C_3]$  and  now the  largest eigenvalue is $\lambda_{max}(z) = z^2 \lambda(z)$
where 
\be 
\lambda(z) =\frac12 (1+ 3z +  \sqrt{ 1+2z + 9 z^2}). 
\ee
In the $z\to 0$ limit,  
\be
\rho(z) = \frac{z}{M}  \frac{d}{dz} \ln \lambda(z) \simeq  \frac{1}{2} + \frac{z}{2}  +{\cal O}(z^2).
\ee
Thus  the critical density  $\rho_c=\frac{1}{2}$  matches  with   the  generic result (\ref{eq:rho_c_evn}) obtained 
for even $M.$ More over in the critical regime,  we have  $z\propto (\rho-\rho_c).$  

The expressions for the  eigenvectors  are  lengthy (omitted here), but the product of  the $\nu_1$ element of 
$|\psi\ra$ and  $(\nu+1)_1$ element  of  $\la\psi|$  is
\be
\la \nu_1|\psi\ra\la\psi|(\nu+1)_1\ra= \frac{1}{ 2 \sqrt{1 + 2 z + 9 z^2}}.
\ee
Using  this   in  Eq. (\ref{eq:master_rhoa}),   in the critical limit we get, 
\be 
\rho_a^*(z) \simeq \frac{z^2}{2} +  {\cal O}(z^2)
\ee
Thus, in the critical regime,  the  order parameter   for $M=4$   behaves as $ \rho_a(z) \sim (\rho-\rho_c)^\beta,$
with $\beta=2.$   For higher $M,$   extracting the order parameter exponent  analytically  using  Eq.
(\ref{eq:master_rhoa}) is   difficult; for  even $M>4$ we proceed  to get an estimate   from
Eq. (\ref{eq:rhoa_evn}). 

In the $z\to 0$ limit,  
\bea
C(z)^L  = z^{\nu L}[ C_\nu^L  + z \sum_{j=0}^{L-1} C_\nu^j C_{\nu+1}  C_\nu^{L-1-j} + {\cal O}(z^2)] \n
\eea
Here,  
from Eq. (\ref{eq:CnRep})we have
\be
C_\nu  =  \Pi_\nu T = (|\nu_1\ra\la\nu_1|  +    |\nu_2\ra\la\nu_2|)T,
\ee
which  has the following properties, 
\bea
&&C_\nu^2=|\nu_1\ra\la\nu_2|T  +    |\nu_2\ra\la\nu_1|T \cr 
&&C_\nu^{2l} = C_\nu^2; ~ C_\nu^{2l+1} = C_\nu \cr
&&  Tr[C_\nu]=0; ~ Tr[C_\nu^2]=2\cr
&&C_\nu  |\nu_1\ra =  |\nu_2\ra; ~~ C_\nu  |\nu_2\ra =  |\nu_1\ra;  
\eea
where $l$ is a positive integer. The proofs   of above relations  are  straight forward,   if  we use the 
facts $\la \nu_k|T |\nu_{k'}\ra = 1- \delta_{k,k'}.$ We  proceed  further   considering  the  system  size to be  $L=2l;$ thus,   to leading order in $z$ 
\be
Tr[C(z)^L] = z^{\nu L} [ Tr[C_\nu^2]  + {\cal O} (z)]  =z^{\nu L}[ 2 + {\cal O} (z)] \label{eq:Z_evn}
\ee
and 
\bea
&&\frac{\la \nu_1| C(z)^{L-2} |(\nu+1)_1\ra}{ z^{\nu (L-2)}} =  \la \nu_1| C_\nu^{L-2} |(\nu+1)_1\ra \cr
&&~~~+ z  \sum_{j=0}^{L-3} \la \nu_1| C_\nu^j C_{\nu+1}  C_\nu^{L-3-j}  |(\nu+1)_1\ra)  + {\cal O} (z^2).\label{eq:C1_evn}
\eea
Now, $\la \nu_1| C_\nu^{L-2} |(\nu+1)_1\ra = \la \nu_1|C_\nu^2|(\nu+1)_1\ra=0$ and we are  left with  
${\cal O}(z)$ term of  Eq. (\ref{eq:C1_evn}). In the  sum, all  the matrix product terms that  ends  with $C_\nu$ would vanish,  because   $C_\nu |(\nu+1)_1\ra = \Pi_\nu T |(\nu+1)_1\ra=0$ as, for $M>6$  the rung $(\nu+1)_1$ can not   be a  neighbor   of  any  of the rungs in the $\nu$-particle sector.
So, the  only  surviving term  in the  sum is  
\bea 
\la \nu_1|C_\nu^{L-3} C_{\nu+1}|(\nu+1)_1\ra= \la \nu_1|C_\nu C_{\nu+1}|(\nu+1)_1\ra=1  \n
\eea
Finally, to  the lowest order in  $z,$   Eq. (\ref{eq:C1_evn})  gives
\be
\la \nu_1| C(z)^{L-2} |(\nu+1)_1\ra =  z^{\nu (L-2)} z,  
\ee
Using  this  and Eq. (\ref{eq:Z_evn}) in  Eq. (\ref{eq:rhoa_evn}) we   obtain, 
\be 
\rho_a^* \simeq z^3,~{\rm  for ~ even~}~  M \ge 6.
\ee

 \begin{figure}\vspace*{.2 cm}
\begin{center}
\includegraphics[width=7 cm]{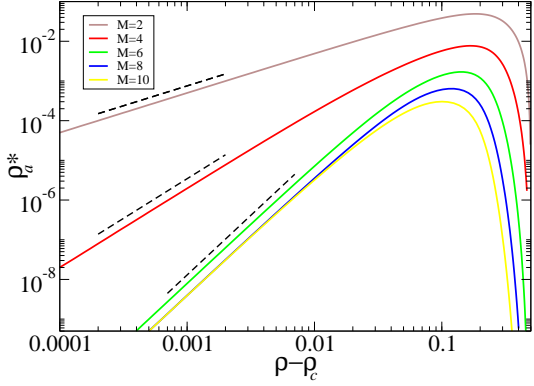}
\caption{(Color online)  Log scale plot of  $\rho_a^*$  shown in  Fig. \ref{fig:evnChain}for  $M=2,4,6,8$ s as a function   of $\rho-\rho_c$  shows  that  $\beta=1,2$ for $M=2,4$  and  $\beta=3$  for even $M>4.$   Lines with slope $1,2,3$  are  shown   in dashed line for comparison.
}
\label{fig:evnC_beta}
\end{center}
\end{figure}

To find the  order parameter  exponent we need to know  the behaviour of $\rho(z)$  at the critical point. As 
$z\to0,$ 
\bea
\rho(z)  &=& \frac{Tr[DC(z)^L]}{Tr[C(z])^L}   \simeq  \frac1{2}  Tr[D C_\nu^L] \cr
&+&  \frac z2 \sum_{j=0}^{L-1} Tr[ D C_\nu^j C_{\nu+1}  C_\nu^{L-1-j}]= \frac12 + \alpha z,     
\eea
where   $\alpha$  is a  positive constant; the proof follows.   Since   each term in the sum is  non-negative,  
the   sum is  larger than  one  specific term say $j=0.$ Again,  since 
$D$ is a  projector   for all  the rungs that  contains $1$ at  first position, 
one of them  $(1,1,0,1,0,1,\dots)$  denoted   by  $\la (\nu+1)_3|$ (which  is rung  $\nu_2$, as in Eq. (\ref{eq:TWO}), with  one extra  particle  at  first position)
gives  $\la\nu_1|T|(\nu+1)_3\ra =1.$  We  obtain an  inequality,  
\bea
&&Tr[ D C_{\nu+1}C_\nu] \ge   \la (\nu+1)_3|C_{\nu+1}C_\nu|(\nu+1)_3\ra \cr
&&= \la (\nu+1)_3|T|\nu_1\ra \la \nu_1|T |(\nu+1)_3\ra  =1.
\eea
This  proves   that   the  constant  $A$ is   nonzero,   and  thus   in the critical regime, 
$ z \propto (\rho-\rho_c)$  and $\rho_a*$  for even $M\ge 6$    is   proportional to   $(\rho-\rho_c)^\beta$
with $\beta=3.$ 

To summarize,  when  the density  approach the  critical value $\rho \to \rho_c,$ 
$\rho_a^*$ and thus        the  order parameter $\rho_a$  behave as,
\bea
\rho_a  
\simeq \frac{1}{2} \left\{ \begin{array}{ll}(\rho-\rho_c)^2 & M=4 \\ (\rho-\rho_c)^3 &  M=6,8,\dots\end{array}\right.
\eea
Thus,   the order parameter  exponent for even  $M$-chain system   is 
\bea
\beta =  \left\{ \begin{array}{ll}1 &{\rm for}~ M=2 ~{\rm (ladder)} \\ 2 & {\rm for}~M=4\\3 & {\rm for}~ M=6,8,\dots\end{array}\right. \label{eq:evn_beta}
\eea
In Fig.  \ref{fig:evnChain} we have   shown the plot of $\rho_a^*$  as a function of $\rho$ for $M=4,6,8,10$ calculated   using    the transfer matrix   formulation (solid line)  and  compared it with the same 
obtained  from  the  Monte-Carlo simulation of the   $M$-chain CLG model, with chain length  $L=1000.$
They clearly indicate that  the absorbing transition occurs at $\rho_c=\frac12.$  In Fig.  \ref{fig:evnC_beta}
the same data, $\rho_a^*$   is plotted   against  $\rho-\rho_c$  in log-scale to obtain the    order parameter exponent  $\beta,$  which agrees  with  Eq. (\ref{eq:evn_beta}).
\section{Conclusion}
\label{sec:conclusion} 
In  this  article we study  the conserved lattice gas  model  on a   multi-chain system, 
where  particles having  exactly one  vacant neighbor  are considered active, and they  are  
allowed to  hop deterministically to  the only vacant neighbor they have.  For   single chain, this 
model  reduces  the  usual  CLG  model in  1D, exhibiting  a nonequilibrium phase transition   from  an active 
to an absorbing  state  when    the density of the system    fall below  a  critical  value  $\rho_c=\frac12;$
the   critical  behavior here  is   rather  trivial, having integer exponents $\beta=1=\nu=\eta.$ 
A two chain   conserved lattice gas model has    been studied earlier \cite{B16}, 
where particles having  at least one   occupied neighbor and one vacant neigbor are    considered  active; absorbing transition in these models  turns out to be in  the directed percolation (DP) universality class,   conjectured  as the  most robust universality class  of absorbing transition.  Since the ladder in the thermodynamic limit     can  be 
considered as a  one  dimensional system,    the change   of universality class  from  1DCLG  to DP 
was rather surprising.  A   possible reason    for  the   flow   to  DP-class  is the stochasticity:  particles 
having  exactly  one  occupied neighbor   must   choose  one  of  the other  {\it two} neighbors (which 
are  vacant)   as the target site and  hop  there.   If  stochastic   particle transfer is a   relevant perturbation, we   should  retain  1DCLG  universality when this  stochasticity is ceased and  
 hopping the dynamics   is restricted  to be deterministic.  Keeping this view in mind we study 
a   restricted   CLG dynamics on a  ladder  (section  \ref{sec:model}) and   indeed,  the APT  
turned out to be in 1DCLG  class. 

It is  natural  to expect that   this scenario    must   prevail  for any  multi-chain $M\times L$ system, 
as  in the  thermodynamic  limit    $L \to\infty$  (keeping   $M$ fixed)    the system is 
effectively  one dimensional. This  is indeed the  case  when  $M$   is   an odd integer  and the APT 
for odd number chains belong to  1DCLG  universality.   The  scenario  is  however  different  
when the number of chains   is   an  even $M \ge 4;$ 
there , the value  of  order parameter  exponent $\beta$  depends on the number of  chains. 
For  $M=2$ (ladder)  the APT  belong to  the 1DCLG universality with $\beta=1$ whereas for  
$M=4$ we get   $\beta=2,$ and for  any  even $M>4$ the order parameter exponent  is $\beta=3.$

\begin{figure}\vspace*{.2 cm}
\begin{center}
\includegraphics[width=8.5 cm]{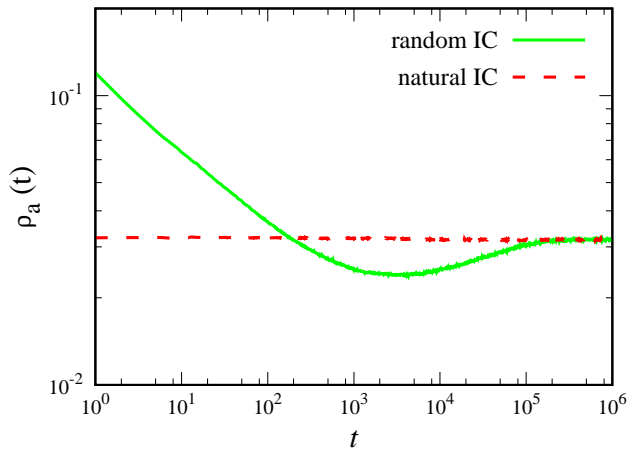}
\caption{(Color online)  The  density  of active particles $\rho_a(t)$  as a function $t$  for $M=2.$ 
The evolution from a   random  initial condition (IC)   exhibit  undershooting and   long-relaxation to the stationary state; both these  ill effects are  avoided if we use  the natural IC.  Here   $L=10^4$ and   
$\rho=0.53.$}
\label{fig:init}
\end{center}
\end{figure}

We calculate  the critical exponents  using a transfer matrix method,   where    the steady state  weight 
can be    written    as  the  trace of a    matrix   string  constructed by  representing   rungs  or vertical supports   of the  $M$-chain systems as matrices.   The number of matrices required for  such a matrix product form is  same as  the number of periodic binary strings which  are devoid of consecutive zeros. This   number, and thus the  dimension  of the transfer  matrix grows  exponentially as $\phi^M$   where $\phi$   is    the golden ratio.  Along with this, the  possible ways  a configuration  can have  local activity  also  grows  quickly and   calculation   of the  the order  parameter $\rho_a,$ which is  the  density of active particles,  becomes practically impossible   as $M$  increases.  However,  
the critical exponent $\beta$ can be  obtained   from   
$\rho_a^*,$  the  steady state average of an  active three-rung configuration   that contains  minimum number of  particles.   We substantiate the calculation $\rho^*$   with  the numerical  
values obtained  from  Monte-Carlo simulation.  

Like  any  other absorbing phase transition  into multiple absorbing  configurations,   Monte-Carlo simulation 
of these models  also  suffers from  the choice of initial condition \cite{sourish} - it is presumed  that  the critical steady state of these systems are  hyperuniform \cite{hexner}  and   the system   takes unusually long time   to  relax and  achieve  that. One must  carefully 
choose initial   conditions  which preserves the  natural correlations of the stationary state.  Again unlike  
1DCLG model ($M=1$)  where  all  supercritical configurations are active,  for  $M\ge 2$ chains  the supercritical 
states    have (i)  absorbing configurations  in supercritical region  and (ii)  active configurations which are  dynamically  inaccessible. For example, when $M$ is  even,  there are   only two configurations at  $\rho_c$  which 
are devoid of CZs;  since one of  the sub-lattice   is completely occupied, in this    configuration 
each particle  have  exactly  four vacant neighbors and  one can   create absorbing configurations  with higher density   by adding  additional particles,  keeping   two neighbors of
every particle vacant.  Also, the  dynamically inaccessible  active configurations  
are   not so uncommon; some  examples  $M=2,3$ are, 
\be
\left(\bm \dots 01\hat1010\dots \\\dots101101\dots\em\right); ~~
\left(\bm \dots10111\dots\\\dots01\hat101\dots\\\dots11110\dots\em\right).
\ee
To avoid  both kind  of configurations in numerical simulations  we  start with   an initial  state that  
contains the rung $|M_1\ra$  which  is fully occupied and the rungs  which  have minimum  number 
$\nu = \lfloor\frac{M+1}2\rfloor;$ of course,  care must be taken so that  the initial   configuration 
is devoid of  CZs. The conserved density of the system $\rho= \zeta  + (1-\zeta) \frac \nu M$  can be 
tuned  by  changing the  number of  $|M_1\ra$  rungs  $\zeta L.$  In Fig. \ref{fig:init} we  plot   $\rho_a(t)$ 
as  a  function of $t$  for density $\rho=0.53$ and $L=10^{4}$ staring from  a  random  initial condition (IC)
(where  $\rho L$  particles  are  placed  at  randomly chosen sites, avoiding multiple occupancy) and 
natural IC, created  from  the  the  rungs $\{ \nu_k\}$  of $\nu$-particle sector and the rung 
$\{M_1\}.$  Clearly  the   random IC  takes  long time to relax  and  produce  under-shooting, whereas the 
natural IC  relaxes  very fast. 

In  the  calculation of the  partition function, however,   we have  summed over all configurations which are devoid of CZs, without avoiding  (i)  absorbing configurations  with $\rho>\rho_c$  and   (ii)  the dynamically 
inaccessible active configurations.  We  presume that  at any supercritical density, the   fraction  of  
such configurations in comparison to the total  number  of configurations devoid of  CZs   
vanishes in the thermodynamic limit.  This assumption  must be true as, for any  $M,$  as 
$\rho_a^*$  obtained  from the numerical simulations  match  with the  analytical results 
obtained using transfer matrix and the partition function; a proof, though  desirable, is  missing.

It  is rather surprising that  the critical  exponents  of these  class  of models depend on the  geometry of the lattice. 
For  even $M,$   system with two or  four chains   for which we get  $\beta=1,2$  respectively, 
may be considered as the   finite  size effect, though unusual. The    most surprising  point  
is the  large $M$  limit, where  $\beta$  explicitly depends on whether  $M$ is  odd  or even; 
in this  case   $M\to \infty$   limit  is nontrivial.  It remains to see, what is the 
critical behaviour of the  restricted CLG model  in   two dimension.


\begin{thebibliography}{50}
\bibitem{footnote1} This is  certainly    
true   when   $\rho> \frac12;$ in that  case  there  might be  consecutive $0$s in the initial configuration, but   eventually 
the  activity  reach   those  regions  and  destroy  them.   However for $\rho<\frac12,$  there  are no configurations  
which  are  devoid of  consecutive $0$s.
\end{thebibliography}
\end{document}